\documentclass[fleqn,usenatbib]{mnras}

\usepackage{newtxtext,newtxmath}

\usepackage[T1]{fontenc}

\DeclareRobustCommand{\VAN}[3]{#2}
\let\VANthebibliography\thebibliography
\def\thebibliography{\DeclareRobustCommand{\VAN}[3]{##3}\VANthebibliography}

\usepackage{graphicx}
\usepackage{hyperref}
\usepackage{float}
\usepackage{threeparttable}
\allowdisplaybreaks

\title[Chemical homogeneity of birth clusters]{Testing the chemical homogeneity of chemically tagged dissolved birth clusters}

\author[C. M. Cheng et al.]{
Chloe M. Cheng,$^{1}$\thanks{E-mail: cmtcheng@uwaterloo.ca (CMC)}
Natalie Price-Jones,$^{1, 2}$
and Jo Bovy$^{1,2}$
\\
$^{1}$Department of Astronomy and Astrophysics, University of Toronto, 50 St George Street, Toronto ON, M5S 3H4, Canada\\
$^{2}$Dunlap Institute for Astronomy and Astrophysics, University of Toronto, 50 St George Street, Toronto ON, M5S 3H4, Canada
}

\date{Accepted 2021 July 10. Received 2021 June 16; in original form 2020 October 19}

\pubyear{2021}

\begin{document}
\label{firstpage}
\pagerange{\pageref{firstpage}--\pageref{lastpage}}
\maketitle

\begin{abstract}
Chemically tagging stars back to common formation sites in the Milky Way and establishing a high level of chemical homogeneity in these chemically tagged birth clusters is crucial for understanding the chemical and dynamical history of the Galactic disc.  We constrain the intrinsic abundance scatter in 17 newly chemically tagged dissolved birth clusters found in the APOGEE survey by modelling APOGEE spectra as a one-dimensional function of initial stellar mass, performing forward modelling of the observed stellar spectra, then comparing the data and simulations using Approximate Bayesian Computation.  We exemplify this method with the well-known open clusters M67, NGC 6819, and NGC 6791.  We study 15 elements measured by APOGEE and find that, in general, we are able to obtain very strong constraints on the intrinsic abundance scatter of most elements in the chemically tagged birth clusters, with upper limits of $\lesssim 0.02$ dex for C, $\lesssim 0.03$ dex for O, Mn, and Fe, $\lesssim 0.04$ dex for Si and Ni, and $\lesssim 0.05$ dex for N, Mg, and Ca.  While we find some evidence for a small amount of chemical inhomogeneity in the remaining elements (i.e. Na, Al, S, K, Ti, and V), we are still able to obtain similar or stronger limits compared to those found for open clusters, consistent with previous findings.  By strongly constraining their level of chemical homogeneity, we can strengthen the statement that these groups of stars represent birth clusters, with promising implications for future chemical tagging studies.  
\end{abstract}




\section{Introduction}\label{sec:introduction}
Uncovering the history of the Galactic disc is a fundamental problem in astrophysics, and understanding its formation as a principal stellar component of the Milky Way is a crucial goal in the theory of galaxy formation.  An intimate knowledge of the disc's chemical and dynamical history can allow us to understand other important processes in stellar and galaxy formation, such as chemical enrichment and star formation history, at a high level of detail.  This will give us greater insight into the Galaxy's present properties and future evolution, and knowing about our own Galaxy will help us understand the wider Universe.  However, it is difficult to achieve this understanding of the formation history of the disc as its stars, gas, and dust are highly dissipated compared to other stellar structures, such as the halo or bulge.  Consequently, much of its chemical and dynamical information has been lost \citep{Freeman_2002, bland_hawthorn_2010, rix_bovy_disc}.

As such, a major goal of near-field cosmology is to reconstruct components of the disc that were gravitationally associated before this dissipation occurred.  A specific example is the reconstruction of birth clusters, the progenitors of star clusters.  While the formation mechanism of star clusters is still an open question (i.e. \citealt{Krumholz_Mckee, Ward}), a widely accepted model is that of giant molecular clouds (GMCs).  In this model, star clusters are born from GMCs in strongly self-gravitating cores of molecular gas, and are thought to emerge as birth clusters in the collapse of their progenitor clouds.  The majority of birth clusters are subsequently dispersed into the disc over a relatively short period of time, with very few remaining gravitationally bound as the star clusters we see today (i.e. \citealt{Lada_Lada, bland_hawthorn_2010, Krumholz_Mckee_BH}).  Thus, the ability to reconstruct many birth clusters over the entire age of the Galaxy would illuminate key information about the chemical and dynamical evolution of the disc.  For example, this would allow for the determination of accurate ages for star clusters, the reconstruction of the stellar cluster mass function (i.e. the relative distribution of initial star cluster masses), and the constraining of radial migration models, to track the manner in which stars have been displaced from their initial radial positions due to interactions with the spiral and bar structures \citep{Ting__2015,natalie}.

A technique by which birth clusters can potentially be reconstructed is that of chemical tagging, which makes use of the immense amount of historical information available in the chemistry of stars.  This technique can be used to attempt to identify birth clusters by associating (or tagging) a large sample of stars with their precise time and site of formation.  In other words, birth clusters can be identified by grouping stars with similar chemistry, in the hopes that these associations represent common formation sites \citep{Freeman_2002, Feltzing_Chiba, ting_chemtag}.

However, the ability to use chemical tagging techniques presupposes the chemical homogeneity of birth clusters.  Chemical homogeneity requires that stars in the same birth cluster have negligible abundance variations at the time of their formation.  This could be possible through two mechanisms: either the GMC in which a birth cluster's member stars formed was uniformly mixed in several chemical elements before star formation began, or a small number of high-mass stars formed early in the GMC's life and enriched it uniformly \citep{Quillen_2002}.  As such, stars associated with the same birth cluster should have some identical chemical abundances \citep{bland_hawthorn_2010, ting_chemtag}.  Abundance variations among birth cluster members can occur at later stages of stellar evolution due to internal mixing, however this mixing should only affect lighter elements that are subsequently synthesized within the stars, whereas the abundances of heavier elements should remain at or near their initial levels \citep{desilva_2009}.  It is reasonable to expect birth clusters to be relatively homogeneous, since the regions in which they form in GMCs are highly turbulent and well-mixed \citep{mckee_tan}.  Turbulent mixing has been shown to produce homogeneous star clusters from GMCs in simulations by \citet{feng}, but finding birth clusters in our Galaxy via chemical tagging and proving that they are truly homogeneous is still a widely explored area.

Several studies have already been completed on the topic of chemical homogeneity, using various samples of open clusters.  Open clusters are useful for chemical homogeneity studies because they represent a very small proportion of birth clusters that have remained gravitationally bound until the present, and so they serve as a sample of the early Galactic disc \citep{allen2006structure, Feltzing_Chiba}.  If it can be shown that open clusters are chemically homogeneous, then it is likely that dispersed birth clusters are chemically homogeneous as well, paving the way for chemical tagging.    

Some studies show that the level of chemical homogeneity within an open cluster is below that of state-of-the-art measurement uncertainties.  For example, \citet{De_Silva_2006} analysed several heavy elements in dwarf stars in the Hyades and their results implied that there is little to no intrinsic scatter between cluster members.  \citet{De_Silva_2007} analysed several elements in 12 red giant stars of Collinder 261 and estimated the intrinsic scatter to be $< 0.05$ dex, with these high levels of homogeneity indicating the preservation of chemical information in this cluster.  \citet{pancino} analysed three red clump stars in five open clusters and found that their abundance patterns were largely close to solar, with [Fe/H] abundances indicative of chemical homogeneity. 

On the other hand, some studies have found inhomogeneity in open clusters; for example, \citet{liu2} studied several elements in 16 stars in the Hyades and found that the Hyades is chemically inhomogeneous at the $0.02$ dex level.  \citet{spina} studied five stars in the Pleiades and found chemical variations that they attributed to planet engulfment events and the evolution of planetary systems.

Other studies, however, have found that these inhomogeneities are limited to comparisons across evolutionary stage, or are so small as to be below measurement uncertainties.  For example, \citet{Blanco_Cuaresma2015} examined 31 open clusters and found that stars at different stages of stellar evolution are chemically inhomogeneous relative to each other.  This may be the result of processes such as NLTE effects, atomic diffusion, mixing, and systematic biases.  Similarly, \citet{Souto2018} studied several different classes of stars in M67, and while they found chemical homogeneity within each evolutionary stage, they found significant variations in chemical abundances across different stages.  \citet{Souto2019} found significant abundance variations between stars in M67 as a function of stellar mass/position on the Hertzsprung-Russell diagram.  Furthermore, \citet{Liu_2019} concluded that while M67 was likely born homogeneous, certain subsets of its members, such as turn-off stars, may be inhomogeneous today, and as such, turn-off stars are not good candidates for chemical tagging.  These three studies attribute these inhomogeneities to atomic diffusion.  Finally, \citet{casamiquela} examined the open clusters Ruprecht 147, the Hyades, and Prasepe and found certain levels of inhomogeneity in each cluster.  They were unable to distinguish the Hyades and Praesepe from each other as these clusters were found to have the same chemical signatures, but encouragingly they could distinguish Ruprecht 147.  

The results of these studies may be contrasting, but what all of these studies have in common is that they use the traditional method of deriving abundances from stellar spectra and comparing scatter in these derived abundances to the measured uncertainties.  This method can be complicated due to incomplete theoretical models of stellar surfaces and instrumental uncertainties \citep{Ness_2015}.  To avoid this, \citet{Bovy_2016} (B16 hereafter) took the less-traditional route of determining tight constraints on the initial abundance spread in the open clusters M67, NGC 6819, and NGC 2420, using high-resolution stellar spectra.  A combined upper limit of $\lesssim 0.05$ dex at 95 per cent confidence was found for nearly all elements studied, using data from the 12th data release (DR12, \citealt{DR12}) of the Apache Point Observatory Galactic Evolution Experiment (APOGEE, \citealt{Majewski_2017}).  Here, we exemplify this method with some of the same clusters as B16 as well as an additional open cluster, using APOGEE's more recent 14th data release (DR14, \citealt{dr14}).  What makes our study unique, however, is that we utilize this approach for a new assessment of chemical homogeneity in a catalogue of several blindly chemically tagged birth cluster candidates, found in \citet{natalie}.  These clusters were found to be homogeneous, despite the fact that their member stars are located at a wide range of distances across the Galactic disc.  We strive to confirm these results by using an entirely different method from \citet{natalie} to examine the level of homogeneity in this intriguing set of stars.  In particular, we constrain the abundance scatter in these birth clusters using stellar spectra instead of derived abundances, which allows us to report the scatter with a very small uncertainty.  This analysis tests the limits of this method and determines whether it is possible to obtain tight constraints on the intrinsic abundance scatter of the remnants of star clusters that have been scattered across the Galactic disc.  

This paper is organized as follows: in Section~\ref{sec:data}, we review the APOGEE survey from which we obtain the data for this study, describe the samples of open and chemically tagged birth clusters, and outline some of our data reduction processes.  In Section~\ref{sec:methods}, we detail the Approximate Bayesian Computation (ABC) method, describe the one-dimensional initial mass model and fitting procedures that we use to characterize the data, and define the process for generating synthetic spectra.  We also describe the algorithm for inferring abundance scatter using forward simulations of the data and ABC.  We present and discuss our results in Section~\ref{sec:results}.  We discuss the implications of our results for chemical tagging in Section~\ref{sec:discussion} and present our conclusions in Section~\ref{sec:conclusion}.  

\section{Data}\label{sec:data}
\subsection{APOGEE}\label{sec:apogee}
We analyse spectroscopic data from DR14 \citep{dr14} of the Apache Point Observatory Galactic Evolution Experiment (APOGEE, \citealt{Majewski_2017}), a component of the Sloan Digital Sky Survey IV (SDSS-IV, \citealt{blanton}).  APOGEE is a high-resolution ($R = 22,500$) spectroscopic survey, observing the H band ($1.51-1.70\mu \text{m}$) using a 300-fiber spectrograph mounted on the 2.5-m Sloan Foundation telescope \citep{Gunn}, located at the Apache Point Observatory.  For each star cluster that we study, we use \texttt{apStar} spectra that combine the observations from all APOGEE observations (`visits' in APOGEE nomenclature) of a given star \citep{holtzman}. 

\subsubsection{Spectral windows}\label{sec:windows}
Associated with APOGEE is the APOGEE Stellar Parameter and Chemical Abundances Pipeline (\textsc{ASPCAP}, see \citealt{aspcap} for details), a software package for the automated analysis of APOGEE spectra.  We make use of \textsc{ASPCAP} in a variety of ways, but an important aspect to highlight for the context of this work is the spectral windows.  \textsc{ASPCAP} uses a set of numbers as a function of wavelength for each element to weight the contribution of different pixels to the $\Delta\chi^2$ used in fitting for the abundance of a particular element.  Thus, high weight is given to pixels that are very sensitive to the abundance of the element and relatively insensitive to the abundances of other elements.  

Our use of the spectral windows represents a key difference between the method used in our study and that in B16.  Since we examine DR14 spectra, we use the DR14 windows, which contain several changes in the line list compared to DR12.  For example, several new lines were added from NIST and other literature publications, and hyperfine splitting was included for some elements \citep{lines}.  

We also make modifications to the weights we use. For C, N, and Fe, we remove several of the smallest non-zero weights from consideration.  This is because, between DR12 and DR14, the number of pixels with non-zero weights available for these elements increases by an order of magnitude, due to the additions to the line list described above (the numbers of non-zero weights for the remaining elements are of similar sizes between DR12 and DR14).  For example, the number of non-zero weights for C increases from 513 in DR12 to 3,800 in DR14, which is concerning since this implies that more than half of the pixels in APOGEE's wavelength range contribute a signal to C.  This means that analysing C using its full set of windows will result in a small amount of signal being picked up from other elements since, despite the efforts of ASPCAP to subtract the contribution of other elements at the same pixel, some contamination still occurs.  The increased number of lines also has the effect of increasing our run times significantly.  A large number of these non-zero weights are extremely small in value, which means they are unlikely to contribute significantly to our analysis.  To mitigate these problems, we make a cut on the DR14 windows for C, N, and Fe by removing the non-zero weights in DR14 that are smaller than the smallest 70\% of the non-zero DR12 weights.  This allows us to use the non-zero weights in DR14 that correspond to the strongest non-zero weights in DR12, keeping our method as similar as possible to B16.  We tested various percentages of the smallest lines to cut and found that 70\% provides us with more reasonable run times and numbers of non-zero weights for C, N, and Fe that are of comparable size to the corresponding set in DR12.  

We make use of 15 elements in APOGEE: C, N, O, Na, Mg, Al, Si, S, K, Ca, Ti, V, Mn, Fe, and Ni.

\subsection{The stellar sample}\label{sec:open_clusters}
\subsubsection{Generalities}
We exemplify our method with three open clusters and use it to analyse 17 chemically tagged birth clusters, all with large numbers of red giant stars ($\geq 10$, consistent with B16).  We analyse giant stars because we only model differences in the chemical absorption line profiles that occur due to temperature and abundance variations.  Additionally, giant stars have negligible initial rotation speeds and inclination angles, which can give rise to additional line variations \citep{Gray, Medeiros}, and they likely represent the initial chemical composition of the cluster to which they belong (see B16 for details).  

To be consistent with the method used in \citet{natalie}, we derive the stellar properties of all stars via \textsc{astroNN}, a neural network designed for high-resolution spectroscopy \citep{astronn}.  We limit our stars to be in the effective surface temperature range $4000\text{ K}\leq T_{\text{eff}} \leq 5000\text{ K}$, since this contains most of the member stars of the clusters we study, and since spectral modelling becomes more uncertain beyond $T_{\text{eff}} < 4000\text{ K}$ (B16).  All clusters and their respective numbers of red giant members between the $T_{\text{eff}}$ limits are listed in Table~\ref{tab:clusters}.  

We obtain our cluster members from two sources.  The example open cluster sample comes from the Open Cluster Chemical Analysis and Mapping (OCCAM, \citealt{Donor_2018}) survey catalogue.  The candidate birth cluster sample comes from a catalogue of blindly chemically tagged stars from across the Galactic disc, found in \citet{natalie} to make up chemically homogeneous groups.  Here, we will give a brief summary of each catalogue and how the membership of their clusters was determined.

\subsubsection{OCCAM}\label{sec:occam}
OCCAM is a comprehensive spectroscopic data set containing information about 19 open clusters in the Milky Way, including their dynamical and chemical parameters and stellar membership.  OCCAM has assigned membership to open clusters in two ways.  First, known members of open clusters observed for APOGEE calibration purposes were chosen, where stars with previously measured abundances and/or high-quality membership studies based on radial velocity were targeted specifically.  Secondly, `likely' cluster targets were chosen on the basis of their location in the colour-magnitude diagram of their parent open clusters.  OCCAM then further constrained membership using Gaia proper motions \citep{gaia}.  The final OCCAM sample contains giant and potential dwarf members.  In this study, we only make use of stars that are flagged as giants due to the considerations we discuss above.  See \citet{Donor_2018} for more details on OCCAM and the membership selection process.  

We primarily use the open clusters as an example of our method for constraining cluster homogeneity. We apply our method to three open clusters using data from the OCCAM survey, including M67 and NGC 6819, which were previously analysed in B16, as well as NGC 6791.  We choose these clusters from the larger OCCAM sample since they have the required number of red giant member stars ($\geq 10$) after the $T_{\text{eff}}$ cuts.  The OCCAM clusters and their numbers of members are reported in Table~\ref{tab:clusters}.

\subsubsection{Blindly chemically tagged birth clusters}\label{sec:pj_clusters}
In \citet{natalie}, the Density-Based Spatial Clustering Applications with Noise algorithm (DBSCAN, \citealt{dbscan}) was applied to a subset of \textsc{astroNN} abundances derived from APOGEE spectroscopic data.  DBSCAN identifies groups as overdensities in chemical space; specifically, it evaluates the chemical space around each star and assigns these stars to groups once it is determined whether there is an adequate number of surrounding stars to classify the region as high-density.  This algorithm was used to blindly chemically tag an eight-dimensional chemical space in APOGEE, resulting in a catalogue of 21 chemically tagged groups with more than 15 members.  These group members are scattered across the entire sample of stars in APOGEE and were found to be chemically homogeneous in each of the 8 abundances used for chemical tagging, as well as in seven other well-measured \textsc{astroNN} abundances.  Members of the groups were also found to be consistent with sharing the same ages.  Additionally, it was found that simulated disrupted open clusters demonstrated orbital action spreads qualitatively similar to those of the chemically tagged groups.  These characteristics led to the belief that these groups represent stellar birth clusters, and studying them could allow us to explore the early Galactic disc and its subsequent evolution at an unprecedented level of detail.  For more information, see \citet{natalie}.

We analyse 17 of the blindly chemically tagged birth cluster candidates, henceforth referred to as `PJ clusters' or `PJ birth clusters', to examine their homogeneity using an entirely different method from that used in \citet{natalie}, in the hopes of further strengthening the claim that these candidates represent real birth clusters.  The PJ clusters we use and their numbers of members are reported in Table~\ref{tab:clusters}.  We choose these particular clusters from the larger sample of PJ clusters since they are the ones with the required number of red giant member stars ($\geq 10$).

\begin{table*}
    \centering
    \caption{Open and chemically tagged birth clusters and the numbers of their members that we analyse.  For the OCCAM clusters, ages are quoted from \citet{Donor_2018}.  For the PJ clusters, the `Age' column refers to the median cluster ages, where the quoted uncertainties represent the median absolute deviation of ages within the cluster, from \citet{natalie}.  The `Number of Members' column refers to the number of stars with temperatures between $4000\text{ K} \leq T_{\text{eff}} \leq 5000\text{ K}$.  The `Total Members Analysed' column refers to the number of stars on which we actually perform analysis, after removing likely red clump stars identified by \texttt{rcsample}.  The `Likely Red Clump Members' column reports the number of stars identified by \texttt{rcsample} as red clump stars, which are removed from the analysed sample of stars.}
    \label{tab:clusters}
    \textbf{OCCAM open clusters \citep{Donor_2018}} \\
    \begin{threeparttable}
    \begin{tabular}{|c|c|c|c|c|}
        \hline
        Cluster name & Age (Gyr) & Number of Members & Likely Red Clump Members & Total Members Analysed \\
        \hline
        M67 & 3.43 & 22 & 3 & 19 (24\tnote{*} ) \\
        NGC 6819 & 1.62 & 23 & 7 & 16 (30\tnote{*} ) \\
        NGC 6791 & 4.42 & 25 & 2 & 23  \\
        \hline
    \end{tabular}
    \begin{tablenotes}
        \item[*] Analysed in B16
    \end{tablenotes}
    \end{threeparttable} \\
    \vspace{0.5cm} 
    \textbf{Chemically tagged birth clusters \citep{natalie}} \\
    \begin{tabular}{|c|c|c|c|c|}
        \hline 									
Cluster name  	&	Age (Gyr)	&	Number of Members	&	Likely Red Clump Members	&	Total Members analysed	\\
\hline 									
PJ 2	&	$9.5\pm0.6$	&	15	&	0	&	15	\\
PJ 3	&	$4.6\pm0.6$	&	15	&	0	&	15	\\
PJ 4	&	$4.9\pm0.5$	&	15	&	1	&	14	\\
PJ 5	&	$10.1\pm0.4$	&	18	&	0	&	18	\\
PJ 6	&	$4.7\pm0.4$	&	16	&	0	&	16	\\
PJ 9	&	$9.5\pm0.5$	&	19	&	0	&	19	\\
PJ 10	&	$3.1\pm0.5$	&	20	&	2	&	18	\\
PJ 11	&	$4.2\pm0.9$	&	16	&	2	&	14	\\
PJ 12	&	$9.7\pm0.1$	&	16	&	0	&	16	\\
PJ 14	&	$8.5\pm0.3$	&	15	&	0	&	15	\\
PJ 15	&	$7.4\pm0.6$	&	15	&	0	&	15	\\
PJ 16	&	$5.0\pm1.0$	&	17	&	2	&	15	\\
PJ 17	&	$5.7\pm1.0$	&	19	&	1	&	18	\\
PJ 18	&	$7.7\pm0.3$	&	12	&	0	&	12	\\
PJ 19	&	$5.6\pm0.5$	&	19	&	0	&	19	\\
PJ 20	&	$5.9\pm0.7$	&	24	&	0	&	24	\\
PJ 21	&	$8.1\pm0.5$	&	14	&	0	&	14	\\
\hline 																					
    \end{tabular}
\end{table*}

\subsection{Uncertainty and error analysis}\label{sec:uncertainty}
APOGEE spectra have reported individual pixel-level uncertainties which are computed from a noise model that tracks the Poisson photon-counting noise.  The correlations between the uncertainties of adjacent pixels are left unrecorded.  Theoretically, these uncertainties should characterize the noise in the spectra realistically.  However, \citet{Nidever} and B16 tested the precision of these reported uncertainties and found that they are significantly underestimated in large areas of the spectra (likely a result of issues with telluric correction) and also display significant correlations (likely a result of the continuum normalization, with perhaps some contribution from scattered light).  We therefore apply several global corrections to the spectra and their associated uncertainties to account for these issues.

To address the underestimation of uncertainties, we assume that uncertainties which result in S/N $> 200$ are underestimated and adjust these to have S/N $= 200$.

APOGEE has also identified several `bad' pixels in each spectrum due to a variety of instrumental and observational effects, through the \texttt{APOGEE\_PIXMASK} bitmask \citep{holtzman}.  To account for this, we set pixels that have non-zero values in the bitmask to be NaNs, removing them from consideration.  A breakdown of the bitmask is given in table 10 in \cite{holtzman}.  We note that, while B16 only made use of the flags \texttt{BADPIX}, \texttt{CRPIX}, \texttt{SATPIX}, \texttt{UNFIXABLE}, \texttt{BADDARK}, \texttt{BADFLAT}, \texttt{BADERR}, \texttt{NOSKY}, and \texttt{SIG\_SKYLINE}, we use all of the bitmask to be more conservative in our method.  

After the bitmask corrections, we find that several pixels left unflagged still have very large errors, with S/N $\sim 0$.  We remove these from consideration by setting pixels with S/N $< 50$ to be NaNs.

\subsection{Red clump removal}\label{sec:RCs}
In B16, red clump (RC) stars were removed from the analysis of C and N in M67 and all of the elements in NGC 6819, as the inclusion of these stars led to complications in the model.  Similarly, we find that cluster members belonging to the RC can minimally affect our constraints on the abundance scatter by leading to variations in the stellar spectra that are not captured by our model.  Additionally, RC stars have a lower $\log{g}$ than first-ascent red giant branch (RGB) stars at the same $T_{\text{eff}}$ in all elements \citep{red_clump}, leading to spectral fluctuations that also cannot be accounted for in our model (see B16 for details).    

It is furthermore important to apply this correction because of the stellar-mass discrepancy at fixed $T_{\mathrm{eff}}$ between the RGB and the RC.  Specifically, our model can account for mixing with stellar mass through its flexible polynomial model for the flux, but it is important to recall that we use $T_{\text{eff}}$ as a proxy for stellar mass (for details, see Section~\ref{sec:model}).  This proxy is appropriate to use along the upper RGB since the relationship between $T_{\text{eff}}$ and stellar mass here is one-to-one.  However, on the lower RGB, $T_{\text{eff}}$ maps to one branch of mass for RGB stars and another branch of mass for RC stars \citep{Bovy_2014}.

\begin{figure}
    \centering
    \includegraphics[width=\columnwidth]{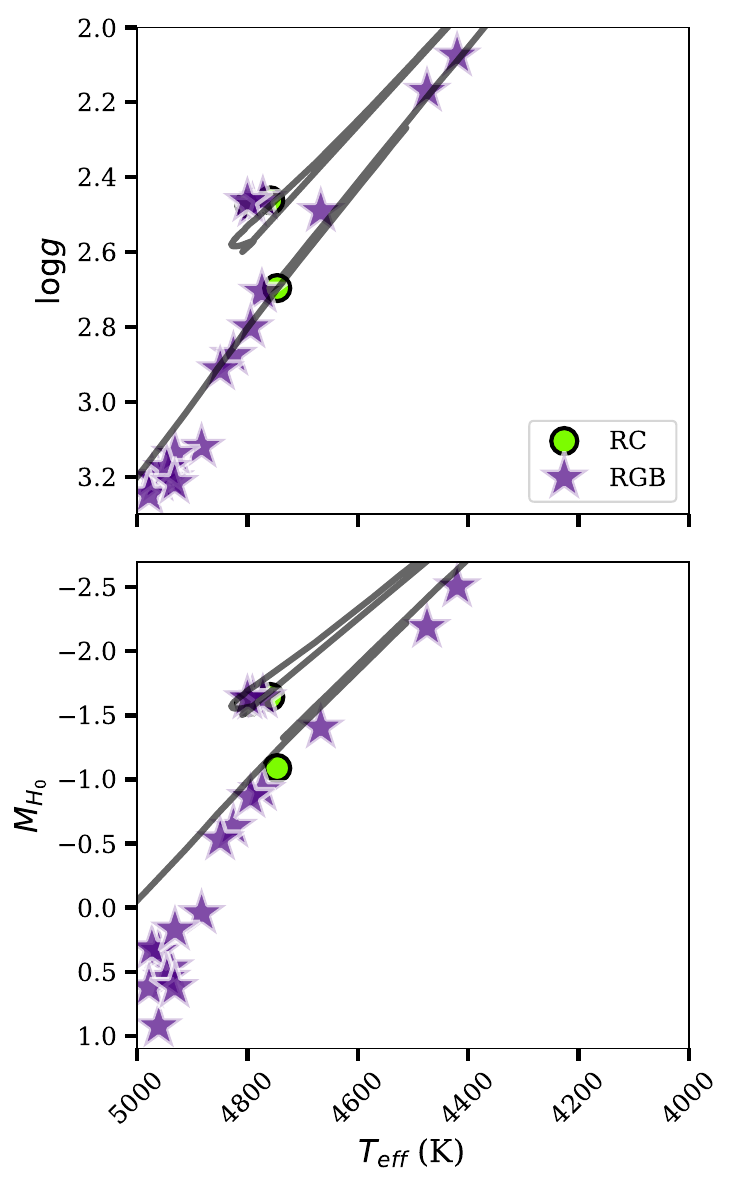}
    \caption{An illustration of the identification of red clump stars via \texttt{rcsample}, using M67 as an example.  The top panel shows $\log{g}$ vs. $T_{\text{eff}}$ and the bottom panel shows absolute $H$-band magnitude ($M_{H_0}$) vs. $T_{\text{eff}}$.  The \texttt{rscample}-identified red clump stars (RC) are shown as green dots and the first-ascent red giant branch stars (RGB) are shown as purple stars.  In each panel, a 2.5 Gyr, [Fe/H] $= 0.0$ PARSEC isochrone is shown as a grey line for comparison \citep{PARSEC}.  The distance to M67 used to compute the absolute magnitudes in the bottom panel is obtained from \citet{M67_dist}.}
    \label{fig:RCs}
\end{figure}

To avoid these issues, we remove all likely RC members from our analysis and use only RGB members.  We identify likely RC members using the \texttt{rcsample} function in the \textsc{apogee Python} package\footnote{\url{https://github.com/jobovy/apogee}~.}, which is part of the procedure to create the APOGEE red clump catalogue\footnote{\url{https://www.sdss.org/dr14/data\_access/value-added-catalogs/?vac\_id=apogee-red-clump-rc-catalog}~.} (see \citealt{Bovy_2014} for details).  While the \texttt{rcsample} function may not identify all RC members perfectly, it is the most consistent and easily reproducible method across all of the clusters we study.  $\log{g}$-$T_{\text{eff}}$ and absolute magnitude-$T_{\text{eff}}$ diagrams are shown in Figure~\ref{fig:RCs} for M67, as an example of the kind of identification that \texttt{rcsample} does.  We note that very few of the PJ clusters have \texttt{rcsample}-identified RC stars, and those that do only have one or two.  As such, the conservative cut of some possible RC stars does not impact our results much, although we perform this cut to maintain consistency with the method from B16, and that which we use for the OCCAM clusters.  

\section{Methods}\label{sec:methods}
This section describes how we model the APOGEE data and the method we use to constrain the level of chemical homogeneity in star clusters.  We make use of the \textsc{apogee Python} package throughout (see above).  

To infer tight constraints on the abundance scatter, we use a method that examines the effect of the abundance scatter on the spectra via forward modelling of the observed stellar spectra, developed in B16.  Specifically, we model the spectra as a one-dimensional function of initial stellar mass.  We also generate synthetic spectra that share the properties of the observed spectra (but with a variety of values of abundance scatter) and model them in the same way.  Finally, we compare the simulations and the data using ABC.  This allows us to determine an upper limit on the intrinsic abundance scatter in each cluster modelled.  

We first describe ABC from a general standpoint (Section~\ref{sec:ABC}), followed by an outline of how we model the data (Section~\ref{sec:model}) and how we generate synthetic spectra (Section~\ref{sec:make_sim}).  We also describe the uncertainty and error corrections that we perform on the observed and simulated data in the context of our method (Section~\ref{sec:summ_uncertainty}).  We identify the summary statistics that we use in the ABC algorithm in Section~\ref{sec:summ_stat}.  Finally, we put all of the pieces together by describing how we use ABC in the context of this study (Section~\ref{sec:ABC_here}).  We also exemplify this method using the open clusters from OCCAM (Section~\ref{sec:occam_results}).  

\subsection{Approximate Bayesian Computation}\label{sec:ABC}
ABC modifies the traditional Bayes' theorem in that it approximates the probability density function (PDF) without first evaluating the likelihood.  Instead, forward simulations of the data are used to approximate the likelihood \citep{ABC}.  Using ABC is advantageous when the likelihood of the data is difficult to compute; in this study, this would involve a full, computable probability model for the residuals from the polynomial fit as a function of $T_{\text{eff}}$ that we perform for different values of intrinsic abundance scatter, which is difficult to generate (B16).

The steps involved in the ABC algorithm are: (a) a set of parameter points is sampled from a prior distribution; (b) for a given sampled parameter point $\theta_p$, a data set $D'$ is simulated using the model $M$; and (c) if $D'$ is significantly different from the observed data $D$ according to a pre-defined threshold, the sample parameter value is rejected.  One or several criteria $\rho(D', D)$ can be defined to represent a distance measure determining how different $D$ and $D'$ are, based on a given metric.  In order for $D'$ to be acceptably close to $D$, $\rho(D', D)\leq\epsilon$, with $\epsilon\geq0$ being the acceptability threshold for `sameness' \citep{ABC}.

As the dimensionality of a data set increases, the probability of being able to generate a simulated data set $D'$ that is close to $D$ decreases.  Thus, a set of lower dimensional summary statistics $S(D)$ that encapsulate the relevant information in $D$ can be used in the algorithm instead of $D$.  Then, the acceptance criterion for the ABC algorithm is defined as $\rho(S(D'), S(D))\leq\epsilon$ \citep{ABC}.  There is inevitably some loss of constraining power in using these summary statistics as they are not sufficient statistics which encode all information related to the inference at hand.  

\subsection{The one-dimensional initial stellar mass model} \label{sec:model}
We model all spectra as a one-dimensional function of initial stellar mass.  This model is described in detail in B16, so we summarize it here and indicate any major deviations from the method.  The framework we use to model spectra is based on the fundamental assumption that there is no initial scatter in the birth abundances of a clusters' member stars; therefore, the only significant difference between individual stars in a cluster is their initial masses.  Thus, we model the stellar spectra near elemental absorption features as a one-dimensional function of initial stellar mass.  Since these measurements are difficult to obtain, we use spectroscopic effective stellar surface temperatures ($T_{\text{eff}}$) of cluster members as a proxy for their initial stellar masses (B16), which are included in APOGEE.  We note that this is a deviation from the method in B16, where photometric $T_{\text{eff}}$ were used.  We use spectroscopic $T_{\text{eff}}$ for the PJ clusters because we do not have a good understanding of their member stars' extinction as a result of their varying distances from us, and so a calculation of the photometric $T_{\text{eff}}$ of these stars might not be accurate.  To maintain consistency, we use spectroscopic $T_{\text{eff}}$ for the OCCAM clusters as well.

For each pixel corresponding to the non-zero weights of each element, we compute a quadratic fit of the spectral flux versus $T_{\text{eff}}$, by applying a least squares fitting routine that accounts for the known Gaussian uncertainties in the data, using the procedure described in \citet{hogg2010data}.  We compute the fit residuals and normalize them by dividing out our corrected pixel-level uncertainties.  We then compute cumulative distributions of the normalized fit residuals using the pixel weights for each element in each cluster individually, as well as for the combined spectra of all cluster members.  For examples of these cumulative distributions, see Section~\ref{sec:occam_results} for the OCCAM open clusters and Section~\ref{sec:results_cumdist} for the PJ birth clusters.  

\subsection{Generating synthetic spectra}\label{sec:make_sim}
To fulfil the forward modelling component (to produce the $D'$) of the ABC algorithm described in Section~\ref{sec:ABC}, we produce several synthetic APOGEE spectra for each cluster by applying the \textsc{Polynomial Spectral modelling} (\textsc{psm}) code by \citet{psm} to the parameters of each star in each cluster.  We simulate chemical abundances as in B16 and feed these into the \textsc{psm} code along with the observed $T_{\text{eff}}$ and $\log{g}$ from APOGEE, and the \textsc{psm} default values of $v_{\text{turb}}$ and $\text{C}_{12}/\text{C}_{13}$ for each cluster ($v_{\text{turb}}$ and $\text{C}_{12}/\text{C}_{13}$ are not provided by APOGEE).

We note that, in contrast to our use of \textsc{psm}, B16 uses \textsc{turbospectrum}.  \textsc{turbospectrum} produces synthetic spectra by directly solving the radiative transfer equation using input data such as opacities \citep{turbospectrum, turbospectrum2}, while \textsc{psm} constructs a second-order polynomial spectral model for the model flux at each wavelength to approximate the spectra. The \textsc{psm} model is fit using a set of $\sim 1,000$ model spectra computed for randomly distributed values in stellar parameters and abundances over the APOGEE giant region. The model spectra are computed using the \textsc{atlas12} and \textsc{synthe} programs \citep{kurucz1, kurucz2, kurucz3}; one advantage of these over the \textsc{turbospectrum} calculations of B16 is that \textsc{atlas12} self-consistently computes the stellar-atmosphere structure for the mix of individual elements, rather than using a pre-computed atmosphere grid in which many element ratios are kept fixed \citep{meszaros_2012}. See \cite{Ting__2016} and \cite{psm} for a detailed explanation of the advantages and disadvantages of \textsc{psm} compared to more traditional codes like \textsc{turbospectrum}, as well as for more information on how model atmospheres are computed.

Because \textsc{psm} only requires the evaluation of X second-order polynomials, where X is the number of wavelength pixels in the spectrum, rather than solving the radiative-transfer equation in full every time, \textsc{psm} is computationally much faster than \textsc{turbospectrum}.  This is part of the reason why we use it.  Specifically, in comparison to the three open clusters with a total of 63 stars examined in B16, we analyse three open clusters and 17 birth clusters, with a total of 335 stars.  Moreover, we simulate each of these stars tens of thousands of times.

Spectra in DR14 and later data releases extend over a slightly wider wavelength range than those in DR12. To account for the fact that the \textsc{psm} code was written for APOGEE DR12, we only consider parts of the observed spectra that lie within the DR12 spectral ranges. The drawback is that some chemical information captured by elemental absorption features located at the edges of the detectors is lost. However, overall, very few absorption lines are cut, leaving us with ample chemical information to analyse.

B16 performed a detailed test of the reported APOGEE uncertainties and determined an empirical model for the spectral errors via repeat observations.  These normalized repeat observation residuals were used as a direct empirical sampling of the noise in the spectra (see B16 for details).  We make use of this repeat observations algorithm to produce $3,150$ repeats residuals for $1,050$ stars in DR14 of APOGEE.  To make our simulated spectra more realistic, we create synthetic noise by drawing a random set of repeats residuals for each star in the observed stellar spectra and multiplying these repeats by the observed spectral errors.  We add these quantities to the simulated spectra to create synthetic noise.  We do this to mitigate the underestimation of the reported APOGEE uncertainties. 

Because the repeats residuals do not necessarily consist of the same stars as the ones we are analysing, they are consequently masked by the \texttt{APOGEE\_PIXMASK} bitmask in different areas, which results in the synthetic spectra being masked in additional places in comparison to the observed spectra.  To account for this, we mask the observed spectra in these places as well.  Additionally, we notice there are some individual repeats with unrealistically large absolute values which affect our results.  We assume this is due to systematic error and avoid these repeats by masking out any pixels in the simulated and observed spectra that correspond to individual randomly selected repeats that are $> 6\sigma$ away from the mean of the repeats.  

\subsection{Summary of uncertainty corrections}\label{sec:summ_uncertainty}
Once we generate the synthetic spectra, we apply the corrections described in Section~\ref{sec:uncertainty} to the observed and simulated spectra.  The order in which we apply the corrections is outlined here:

\begin{enumerate}
    \item For the observed and simulated spectra:
\begin{enumerate}
    \item Adjust underestimated uncertainties with S/N $> 200$ to have S/N $= 200$.
    \item Apply the \texttt{APOGEE\_PIXMASK} bitmask (these bits will be set to NaNs and neglected).
    \item Mask large-error pixels with S/N $< 50$ (these points will be set to NaNs and neglected).
    \item Mask pixels outside of the bounds of the DR12 detectors.
\end{enumerate}
\item For the simulated spectra:
\begin{enumerate}
    \item Produce mock noise to add to the spectra by drawing from the repeats residuals, which will further mask the spectra.
    \item Mask values of the chosen repeats that are $> 6\sigma$ away from the mean. 
\end{enumerate}
\item For the observed spectra:
\begin{enumerate}
    \item Re-mask the observed spectra to match the new masking in the simulated spectra that results from the application of the repeats residuals.
\end{enumerate}
\end{enumerate}

Once we apply these corrections, we fit each observed and simulated spectrum using the one-dimensional initial mass model (Section~\ref{sec:model}), fitting only those pixels that have $\geq 5$ stars remaining after all uncertainty and error considerations have been applied.  If an element has $<$ 5 stars remaining, we neglect that element in our analysis to ensure that the best-fitting parameters can be precisely determined.  To illustrate this with an example, the open cluster NGC 6791 has 23 member stars which have $4000 \text{ K} \leq T_{\text{eff}} \leq 5000 \text{ K}$ and are not marked as red clump stars in \texttt{rcsample}.  We analyse these 23 stars in the cluster overall, however, at least 19 of these stars are set to NaNs after the uncertainty and error corrections in the pixels for the element Na specifically.  As such, we cannot perform significant fits on the pixels of Na for NGC 6791, however, we can still perform fits using all 23 stars in, for example, the element C.   

\subsection{Summary statistics}\label{sec:summ_stat}
We make use of two summary statistics computed using the residuals from the one-dimensional fit to each wavelength pixel: the Kolmogorov-Smirnov distance and a statistic describing the difference between the covariance matrices of the normalized residuals of different pixels for the observed and simulated data (B16). 

The Kolmogorov-Smirnov distance ($D_n$) is defined as the maximum absolute difference between the cumulative distributions of the normalized residuals of the data and the normalized residuals of a simulation.  We compute the covariance matrix summary statistic, $|\Delta \text{Cov}_{\text{ij}}|$, as in Equation (7) in B16.

\begin{table}
    \centering
    \caption{The 95 per cent upper limits on the constrained intrinsic abundance scatter for the 15 APOGEE elements for three open clusters in OCCAM.  A dashed entry indicates that not enough data are available for the element in question after we complete the masking and uncertainty corrections (i.e. $< 5$ pixels are available to fit that element).}
    \begin{tabular}{c|c|c|c|}
\hline							
	&	M67	& 	NGC 6819	&	NGC 6791	\\
\hline							
C	&	0.035	& 	0.041	&	0.046	\\
N	&	0.042	& 	0.071	&	0.069	\\
O	&	0.057	& 	0.08	&	0.083	\\
Na	&	0.092	& 	0.095	&	-	\\
Mg	&	0.032	& 	0.04	&	0.083	\\
Al	&	0.036	& 	0.097	&	0.096	\\
Si	&	0.027	& 	0.034	&	0.085	\\
S 	&	0.074	& 	0.097	&	0.097	\\
K	&	0.072	& 	0.078	&	0.094	\\
Ca	&	0.038	& 	0.035	&	0.096	\\
Ti	&	0.055	& 	0.062	&	0.097	\\
V	&	0.091	& 	0.091	&	0.099	\\
Mn	&	0.025	& 	0.027	&	0.089	\\
Fe	&	0.018	& 	0.031	&	0.08	\\
Ni	&	0.038	& 	0.037	&	0.09	\\
\hline																	
    \end{tabular}
    \label{tab:my_results_occam}
\end{table}

\subsection{The ABC algorithm in the context of this study}\label{sec:ABC_here}
Once we compute the summary statistics, we quantitatively determine the constraints on the intrinsic abundance scatter in our clusters by using the ABC algorithm to construct an approximation of the posterior PDF of the scatter in each element.

\begin{figure*}
    \centering
    \includegraphics[width=\textwidth]{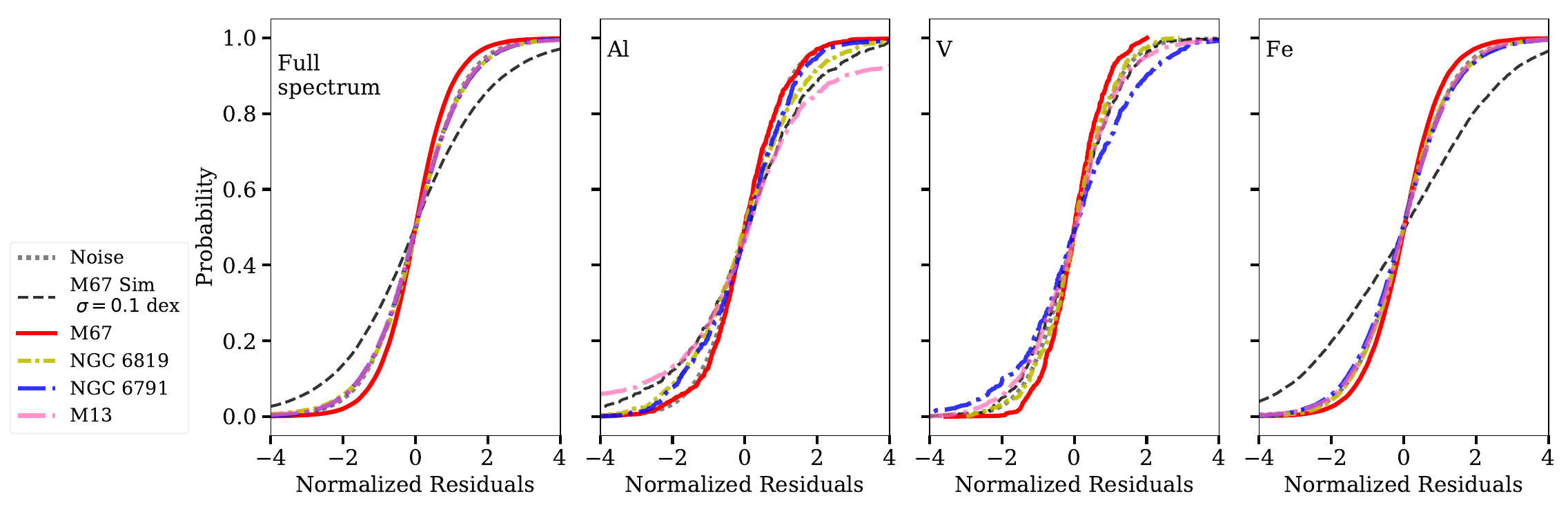}
    \caption{The cumulative distributions of the normalized residuals from the quadratic $T_{\text{eff}}$ fits for the OCCAM open clusters.  The cumulative distribution of all pixels in the spectra is shown in the leftmost panel for each cluster and the subsequent panels display the cumulative distributions for the fit residuals weighted by the pixel weights that emphasize the pixels most affected by Al, V, and Fe, for example (see Section~\ref{sec:windows}).  The cumulative distribution of the repeats residuals (as we discussed in Section~\ref{sec:make_sim}) is represented by the dotted grey curve, corresponding to the noise in the spectra.  Additionally, the cumulative distributions of the normalized residuals for a simulation of M67 with $\sigma_{\text{[X/H]}} = 0.1$ dex for each element X is represented by the dashed black curve; this level of intrinsic abundance scatter would produce much larger residuals than observed for Al and Fe, but is similar to that in V.  We also show the cumulative distribution functions for the same elements in the globular cluster M13 for comparison, which has well-known scatter and anticorrelations in light element abundances (i.e. Al), but relatively little scatter in heavier elements (i.e. V and Fe, \citealt{meszaros_2015}, B16).}
    \label{fig:cdists_occam}
\end{figure*}

For each set of simulations for a particular element, we plot the values of $D_n$ and $|\Delta\text{Cov}_{\text{ij}}|$ on a scatter plot.  Examples of these plots are shown in Section~\ref{sec:results_summstat} below, for the PJ birth clusters.  To determine the upper limits on the intrinsic abundance scatter in the clusters we examine, we select simulations that `match' the data in that they have small values of $|\Delta\text{Cov}_{\text{ij}}|$ and $D_n$.  We then create cumulative posterior distribution functions of the corresponding values of $\sigma_{\text{[X/H]}}$, examples of which are shown in Sections~\ref{sec:occam_results} (for the OCCAM clusters) and \ref{sec:results_constraints} (for the PJ clusters).  From these functions, we determine the upper limits on the intrinsic abundance scatter for each element at a confidence level of 95 per cent.  This allows us to identify the value of the abundance spread that produces simulated spectra that are closest to the observed spectra, thus constraining the abundance spread.

We run simulations until these limits and the distributions of $D_n$ and $|\Delta\text{Cov}_{\text{ij}}|$ appear to have converged by eye ($\sim 10,000$ simulations in total and we use $\sim 700 - 1,000$ simulations that are very close to the data to determine the upper limits on the intrinsic abundance scatter).  Once we have achieved convergence, we report the final upper limits on the intrinsic abundance scatter at a confidence level of 95 per cent.

\subsection{Example of the method with open clusters}\label{sec:occam_results}
To exemplify our method, we run the algorithm on the three open clusters from OCCAM.  We perform the one-dimensional initial stellar mass fitting procedure on the observed spectra, where we see that the distribution of the residuals is consistent with the uncertainty distribution for each pixel of the observed spectra, similar to the results in B16, indicating that this one-dimensional model provides a good fit.  We also perform this fitting procedure on the simulated data sets. For simulations with $\sigma \rightarrow 0.1$ dex, the residuals are much larger than the reported uncertainties, as in B16, implying that the scatter around the quadratic fit is strongly constraining for the intrinsic abundance scatter.

\begin{figure*}
    \centering
    \includegraphics[width=\textwidth]{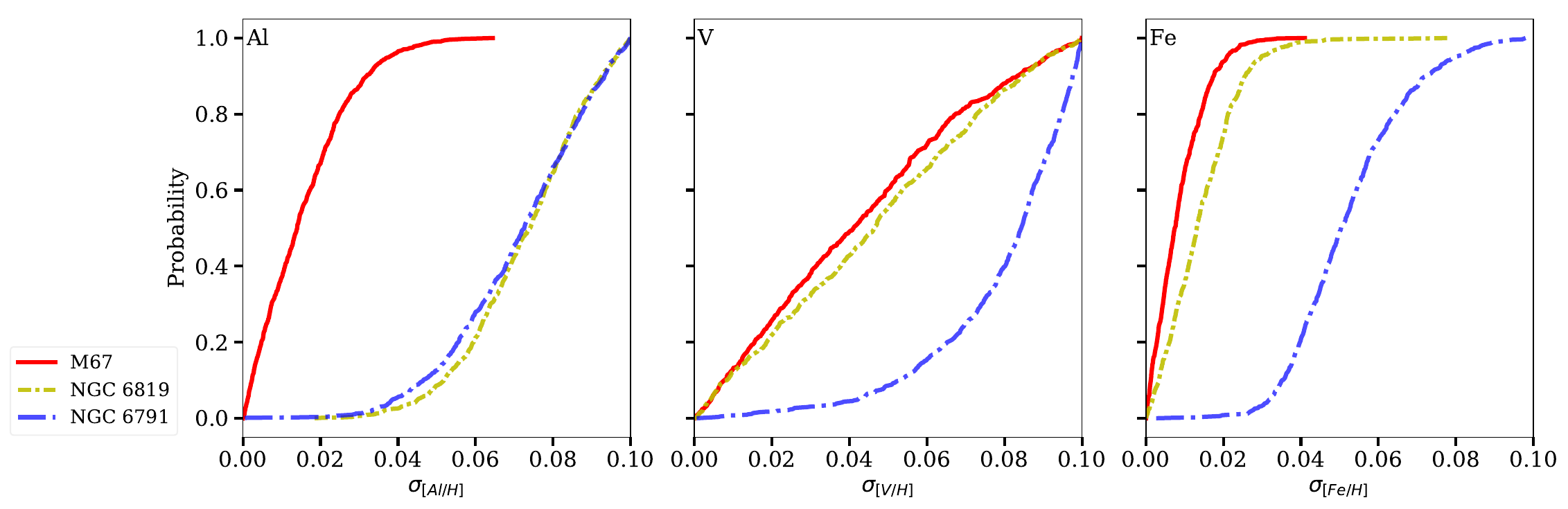}
    \caption{The cumulative posterior distribution functions for the intrinsic abundance scatter in Al, V, and Fe as examples, in each of the three open clusters from OCCAM.  The 95 per cent upper limits for each OCCAM cluster are reported in Table~\ref{tab:my_results_occam}.}
    \label{fig:posterior}
\end{figure*}

We then compute cumulative distributions of the normalized fit residuals.  A comparison of the cumulative distributions for each OCCAM open cluster to the cumulative distributions of the noise and normalized residuals from a simulation of M67 with $\sigma_{\text{[X/H]}} = 0.1$ dex are shown in Figure~\ref{fig:cdists_occam}, for the elements Al, V, and Fe, and for the combined elements in each cluster.  We choose to show this subset of elements as an example of some of the different types of behaviour that we see (although we make similar visualizations for every element in APOGEE in our analysis).  For example, it is clear that the simulation has a much wider distribution than that of the observed spectra for Al and Fe.  However, the simulation has a similar distribution compared to the observed spectra for V.  For Al and Fe, the cumulative distribution of the data is steeper and centred at zero, indicating that the fit residuals lie at or near zero and that the quadratic $T_{\text{eff}}$ model fits the data well.  The behaviours of C, N, O, Mg, Si, Ca, and Mn are similar to Al and Fe and the behaviours of Na, S, K, and Ti are similar to V.

For comparison and to confirm the validity of our method, we perform the same procedure on 46 members of the globular cluster M13, similar to B16.  As discussed in B16, M13 is known to have significant intrinsic scatter in the abundances of its light elements, demonstrating anticorrelations in elements like Mg and Al \citep{meszaros_2015}.  The cumulative distributions of the normalized residuals from the quadratic $T_{\text{eff}}$ fits for M13 are shown in Figure~\ref{fig:cdists_occam}, alongside the OCCAM cumulative distributions.  It is clear that the abundance spread in M13 is much greater than that in M67 for light elements such as Al.  Similar behaviour appears for C, N, O, Na, and Mg, and this behaviour is largely consistent with the findings in \cite{meszaros_2015}, who performed a detailed analysis of several globular clusters including M13.  This indicates that our method is able to recover the expected intrinsic abundance scatter in the light elements of globular clusters like M13.

As an aside, we note that this method cannot be used on field stars.  One of the underlying requirements of this method is that the population being analysed must be a single stellar population, which does not hold in the field.  Thus, in this case, the one-dimensional assumption in the model does not hold.

To quantitatively determine the constraints on the abundance scatter in the open clusters, we apply the ABC algorithm.  Examples of summary statistic scatter plots are shown in Section~\ref{sec:results_summstat} below for some of the PJ clusters and we make similar visualizations in the analysis of the OCCAM open clusters. 

We compute cumulative posterior distribution functions from simulations of each cluster in every element, using the simulations with values of $\sigma_{\text{[X/H]}}$ that are close to the data (captured by threshold boxes similar to those in Section~\ref{sec:results_summstat} below for the PJ clusters).  These cumulative posterior distributions are shown in Figure~\ref{fig:posterior} for Al, V, and Fe in each cluster in OCCAM, as examples.  We show these elements as examples for similar reasons as above.   

\begin{figure*}
    \centering
    \includegraphics[width=\textwidth]{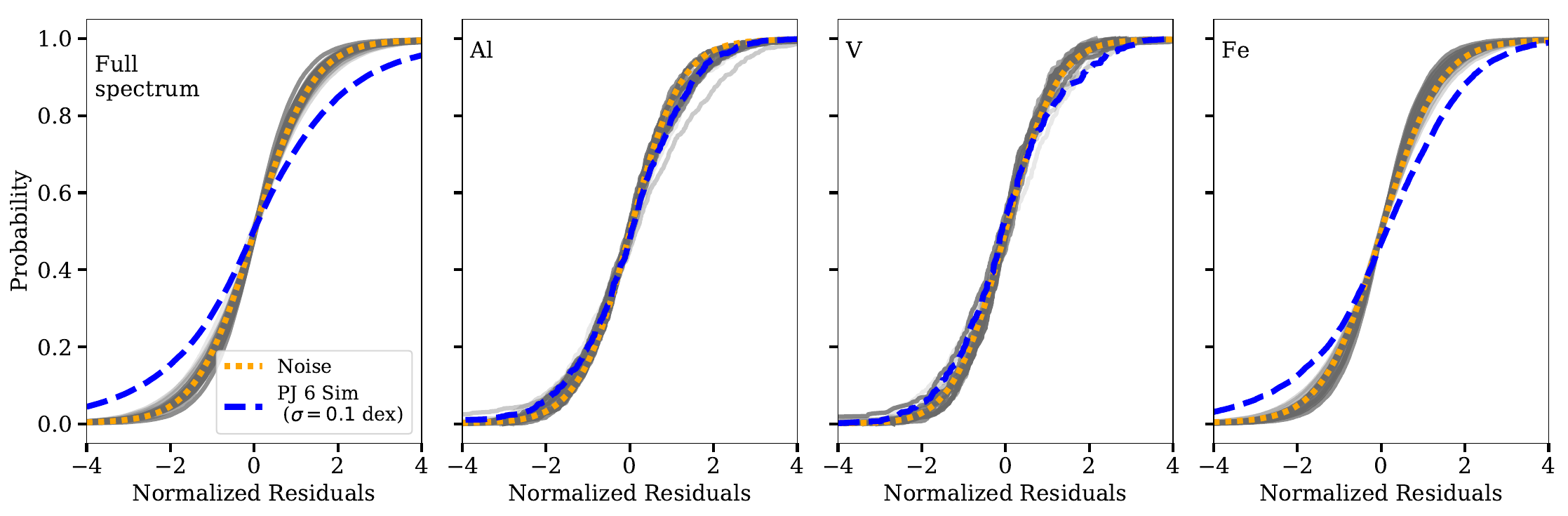}
    \caption{The cumulative distributions of the normalized residuals from the quadratic $T_{\text{eff}}$ fits for the PJ birth clusters.  The cumulative distribution of all pixels in the spectra is shown in the leftmost panel for each cluster and the subsequent panels display the cumulative distributions of the normalized fit residuals weighted by the pixel weights that emphasize the pixels most affected by Al, V, and Fe, respectively.  In each panel, the cumulative distributions of each of the PJ birth clusters are represented by the series of grey curves.  The cumulative distributions of the repeats residuals are represented by the dotted orange curve, corresponding to the noise in the spectra.  Additionally, the cumulative distributions of the normalized residuals for a simulation of PJ 6 with $\sigma_{\text{[X/H]}} = 0.1$ dex for each element X are represented by the dashed blue curve. Overall, the fit residuals for the PJ clusters are consistent with the noise, thus requiring little-to-no intrinsic abundance scatter.}
    \label{fig:cdists_pj}
\end{figure*}

The upper limits on the intrinsic abundance scatter at 95 per cent confidence for each element are reported in Table~\ref{tab:my_results_occam} for the OCCAM open clusters.  We obtain strong constraints ($\lesssim 0.05$ dex at 95 per cent confidence) on most of the elements in M67 (except for O, Na, S, K, Ti, and V) and some of the elements in NGC 6819 (except for N, O, Na, Al, S, K, Ti, and V).  We obtain weak limits on nearly all of the elements in NGC 6791, except for C (although this is not surprising given that NGC 6791 is very metal-rich, \citealt{Donor_2018}).

Comparing the results of our example to those from B16 for M67 and NGC 6819, there are some minor differences in the upper limits on the intrinsic abundance scatter of these open clusters.  The weaker constraints that we obtain for elements such as Na, S, K, Ti, and V are not necessarily surprising, since these elements have very few pixels over which we are able to do our analysis compared to the remaining elements in APOGEE.  This is due to our choice of bitmask, which is more aggressive than that used in B16, and our uncertainty corrections.  Additionally, the abundances for Na, Ti, and V have been cited as being unreliable in APOGEE \citep{Ting__2015}.  This is also reflected in the cumulative distribution (Figure~\ref{fig:cdists_occam}) and cumulative posterior distribution (Figure~\ref{fig:posterior}) for V (those for Na, S, K, and Ti look similar).

The minor differences in the upper limits on the remaining elements can be explained by the differences in the method used between this study and that in B16.  For example, as described in Section~\ref{sec:windows}, there are significant differences between the DR12 windows and the DR14 windows, and a certain amount of contamination from other elements could be affecting some of our results.  Additionally, our more aggressive masking and uncertainty-correction procedures may lead to weaker overall constraints on the intrinsic abundance scatter of open clusters.  As a result of this and our RC star removal, we use fewer stars in our analysis of M67 and NGC 6819 than B16.  We choose to alter the method in these ways to be more conservative, because it is important to be very conservative when we analyse the PJ birth clusters in Section~\ref{sec:results}, since they are still candidate birth clusters and we want to properly strengthen the claim that their member stars were born in the same places.

The cluster membership between this study and B16 is also different, due to variations between the catalogue from \citet{Jo_cat} used to identify open cluster members in B16 and our use of OCCAM \citep{Donor_2018} (i.e. of the 19 members we analyse in M67, 12 of them are analysed in B16 and of the 23 members we analyse in NGC 6819, three of them are analysed in B16).

A final difference between this study and B16 is the way in which we generate our synthetic spectra, which we discussed in Section~\ref{sec:make_sim}.  

Taking all of these differences into account, we are able to achieve relatively similar constraints on the abundance scatters in the open clusters M67 and NGC 6819 when compared to B16, excluding some of those elements that have very few pixels over which we can do our analysis (specifically Na, S, K, and V).  Thus, we have calibrated our method via this example using the open clusters, and can use it to analyse the chemically tagged birth clusters from \citet{natalie}.

\section{Results and discussion}\label{sec:results}
In this section, we present and discuss the results of our analysis of the PJ birth clusters.  In Section~\ref{sec:results_cumdist}, we present the cumulative distributions of the normalized residuals for each of the PJ clusters and discuss their implications.  In Section~\ref{sec:results_summstat}, we show some example summary statistic plots and discuss their different behaviours that indicate chemical homogeneity or lack thereof.  In Section~\ref{sec:results_constraints}, we present our constraints on the intrinsic abundance scatter for each element in each PJ cluster, as well as the combined constraint for each element.  In Section~\ref{sec:results_summaryplts}, we visually summarize our constraints.

\subsection{Cumulative distributions}\label{sec:results_cumdist}
Figure~\ref{fig:cdists_pj} shows a representation of the cumulative distributions from all of the analysed PJ clusters, represented by the series of grey curves, compared to cumulative distributions of the noise (computed from the repeats residuals) and of the normalized fit residuals from a simulation of PJ 6 with an abundance scatter of $\sigma_{\text{[X/H]}} = 0.1$ dex.  The cumulative distribution of all of the pixels in the spectra is shown in the leftmost panel for each cluster and the subsequent panels display the cumulative distributions of the fit residuals for Al, V, and Fe.  We choose to display these particular elements as examples for the same reasons as in Section~\ref{sec:occam_results} for OCCAM.  Similar to Figure~\ref{fig:cdists_occam}, it is clear that the simulation has a wider distribution than that of the observed spectra for Al and Fe (although this does not hold for V, for similar reasons as the OCCAM open clusters), and that the quadratic $T_{\text{eff}}$ model fits the data well.  

\subsection{Summary statistic plots and potential chemical inhomogeneity}\label{sec:results_summstat}
\begin{figure*}
    \centering
    \includegraphics[width=\textwidth]{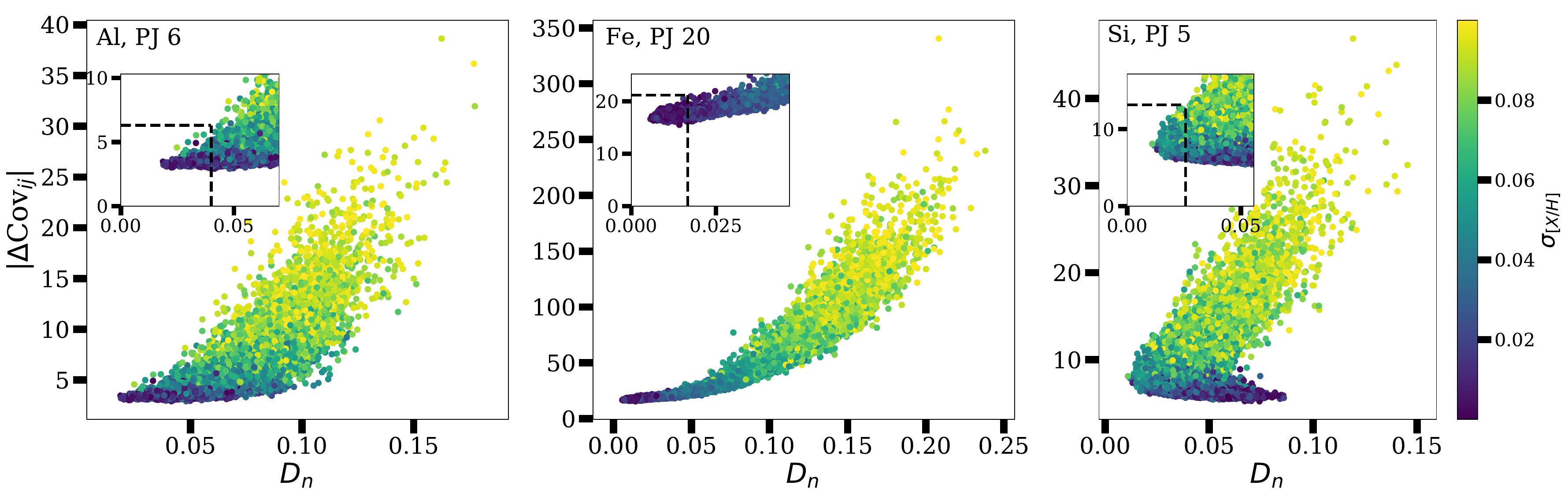}
    \caption{A visualization of the two summary statistics we compute in the ABC algorithm to compare the observed spectra to the simulated spectra, for a sample of the chemically tagged birth clusters that we study.  The left-hand panel displays the simulations with intrinsic scatter in Al in PJ 6, the middle panel displays those for Fe in PJ 20, and the right-hand panel displays those for Si in PJ 5.  The Kolmogorov-Smirnov distance ($D_n$) is shown on the $x$-axis and the covariance matrix statistic ($|\Delta\text{Cov}_{\text{ij}}|$) is shown on the $y$-axis.  To compute these summary statistics, we use the residuals weighted for each respective element in each panel.  The colour bars represent the different values of $\sigma_{\text{[X/H]}}$ that we use for each simulation, where X is Al, Fe, or Si in these examples.  The inset panels display the simulations that are closest to the data (i.e. where both summary statistics are small) and the dashed lines within show the thresholds we use to constrain $\sigma_{\text{[X/H]}}$.}
    \label{fig:summ_stats}
\end{figure*}

Examples of distributions of the summary statistics from the ABC algorithm obtained for each set of simulations are shown in Figure~\ref{fig:summ_stats}, where $D_n$ is shown on the $x$-axis and $|\Delta\text{Cov}_{\text{ij}}|$ is on the $y$-axis.  The colour bar represents the different values of $\sigma_{\text{[X/H]}}$ that we draw for each simulation.  The left-hand panel displays the summary statistics for simulations with intrinsic scatter in Al in PJ 6, the middle panel displays those with intrinsic scatter in Fe in PJ 20, and the right-hand panel displays those with intrinsic scatter in Si in PJ 5.  We show these particular elements as examples of some of the different types of behaviour that we see, but we create visualizations like these for every element in every birth cluster we study.  

The panels for Al in PJ 6 and Fe in PJ 20 in Figure~\ref{fig:summ_stats} represent ideal summary statistic behaviour.  It is important to note that, while most of the elements that have enough pixels for analysis display this behaviour, not all elements in all of the clusters we study are as well-behaved.  Specifically, some elements show a `turning-back' behaviour, which indicates potential intrinsic abundance scatter.  For example, in the panel for Si in PJ 5 in Figure~\ref{fig:summ_stats}, the tail of simulations with $\sigma_{\text{[X/H]}}\rightarrow0$ turns back towards increasing $D_n$, instead of being focused at small $D_n$ and $|\Delta\text{Cov}_{\text{ij}}|$ as in the preceding panels.  Specifically, this plot seems to indicate that there is intrinsic abundance scatter in Si in PJ 5.  Similar behaviour is also seen in some of the other PJ clusters, summarized in Table~\ref{tab:bad_elems}.  However, the behaviour of the majority of the summary statistic plots of the PJ clusters is generally indicative of chemical homogeneity.

\begin{table}
    \centering
    \caption{Elements in specific PJ clusters that show behaviour indicative of intrinsic abundance scatter.  However, overall, the majority of elements in all clusters show behaviour indicative of chemical homogeneity.}
    \begin{tabular}{c|c}
        \hline
        PJ cluster & Elements with scatter \\
        \hline
        PJ 4 & Mg, Al \\
        PJ 5 & All elements \\
        PJ 9 & C, Al, Si, Ni \\
        PJ 10 & Al, Si \\
        PJ 12 & Mn \\
        PJ 14 & O, Fe \\
        PJ 15 & Mg \\
        PJ 16 & Ti, Mn \\
        PJ 17 & O, Mg, Al, Si, Ca, Ti, Mn, Fe \\
        PJ 18 & O, Si, Ca, V, Mn \\
        PJ 19 & O, Mg, Si, Ca, Ti, Mn, Ni \\
        PJ 21 & O, Mg, Al, Si, Ca, Ti, Mn, Ni \\
        \hline
    \end{tabular}
    \label{tab:bad_elems}
\end{table}

It is not necessarily clear why this discrepancy in homogeneity between different elements in different clusters exists, however we will discuss potential reasons here.  One explanation could be that there is some aspect of the spectra we are not taking into account in our modelling of these elements.  There are several factors that could invalidate our assumption that non-elemental influences on the spectra can be modelled as a one-dimensional function of $T_{\text{eff}}$.  This could explain some of the weaker upper limits we obtain and the apparent abundance scatter in some elements.  For example, there may be systematic uncertainties in these clusters that cannot be modelled as a smooth function of initial stellar mass, mixing effects that we are not considering (such as deep mixing, which can change the surface abundances of stars), or other real-world complications that we are not modelling properly in our forward simulations (such as the existence of a binary companion, with mass transfer between binary companions resulting in spectral scatter). Such effects would be difficult to correct by any method (B16).  As another example, stars can have a variety of initial rotation velocities, which can result in differences in the spectra of younger cluster members that cannot be modelled as a function of initial stellar mass \citep{Nielsen}.  Initial velocity differences may also present complications, since they can result in present-day spectral scatter in the case that they brought about different mixing histories in the past.  This will result in altered surface abundances of elements such as C and N \citep{Pinsonneault, Meynet}.  

For the most part, we circumvent these effects by examining giant stars.  For example, cooler stars with deep convective zones have stronger magnetic fields, and so they lose angular momentum faster as a result of stellar winds \citep{Nielsen}, making our use of red giant stars in a temperature range of $4000\text{ K} \leq T_{\text{eff}} \leq 5000\text{ K}$ beneficial.  However, different viewing angles will still result in spectral scatter.  For the method in B16 to work for every stellar type, the $v\sin{i}$ of each star would need to be deduced prior to the performance of the forward simulations.  Again, the fact that we study giant stars here should minimize this effect, since their rotational velocities are small (B16), but perhaps the role of this effect needs to be considered more carefully in order to fully characterize the chemical homogeneity of these clusters.  

On the other hand, it is possible that we are modelling these clusters in a robust way and that they actually display signs of chemical inhomogeneity.  There are many examples of effects that can cause present-day chemical abundances to be inhomogeneous, but sources of initial inhomogeneity are less obvious.  However, perhaps the way in which stars are born can offer some insight into the reasoning for this effect.  As stated in Section~\ref{sec:introduction}, it is thought that most stars are born in birth clusters, which form in the collapse of a GMC, and the two predictions for methods by which initial chemical homogeneity occurs is that the progenitor gas cloud was uniformly mixed in critical elements before stars began to form, or a few high-mass stars may have formed soon after the development of the cloud, enriching the cloud in a uniform way \citep{Quillen_2002}.  Perhaps our results indicate that the formation scenario for star clusters is different than what is commonly assumed (i.e. \citealt{Krumholz_Mckee, Ward}), or perhaps these two predictions need to be re-evaluated.  Further work needs to be done to examine this.  

There are numerous other effects that may result in different levels of chemical homogeneity in different elements, which could explain why some elements suggest intrinsic abundance scatter compared to other elements in the same cluster.  For example, planet formation can affect the chemical composition of stars \citep{Liu_2019}.  Refractory elements in the proto-stellar nebula, for example C, N, O, Si, S, Fe, and Ni \citep{refractory}, which are elements that we analyse, may have been locked up in early-forming terrestrial planets, causing the gas left over for star formation and therefore the gas within the stars themselves to have reduced abundances of these refractory elements.  On the other hand, early planetary infall into stars could increase the abundances of refractory elements in these stars \citep{Liu_2019}.  These selective planetary-forming elements could explain why some elements in a cluster display chemical homogeneity while others do not.  A potential explanation for why chemical homogeneity is more apparent in PJ 6 than PJ 5, for example, is that perhaps planets were more prevalent in PJ 5 before the members of this birth cluster became dispersed across the Galactic disc, relative to PJ 6. Further investigation would need to be done to prove this, for example, looking for a trend in abundance scatter with condensation temperature (i.e. \citealt{tcond1}, \citealt{tcond2}).  

Finally, elements such as Na, Mg, V, and Mn can show a significant gradient in abundance with $T_{\text{eff}}$ and $\log{g}$ due to non-LTE effects, changes of chemical abundances due to stellar evolutionary stage, and analysis systematics such as the blending of absorption lines; elements with a small number of lines, such as Mn, are likely to be affected by this \citep{casamiquela}.  This could be a potential explanation for the inhomogeneity that we see in Mg in PJ 4 or Mn in PJ 18 and PJ 19, for example.  

\subsection{Abundance scatter constraints}\label{sec:results_constraints}
\begin{figure*}
    \centering
    \includegraphics[width=\textwidth]{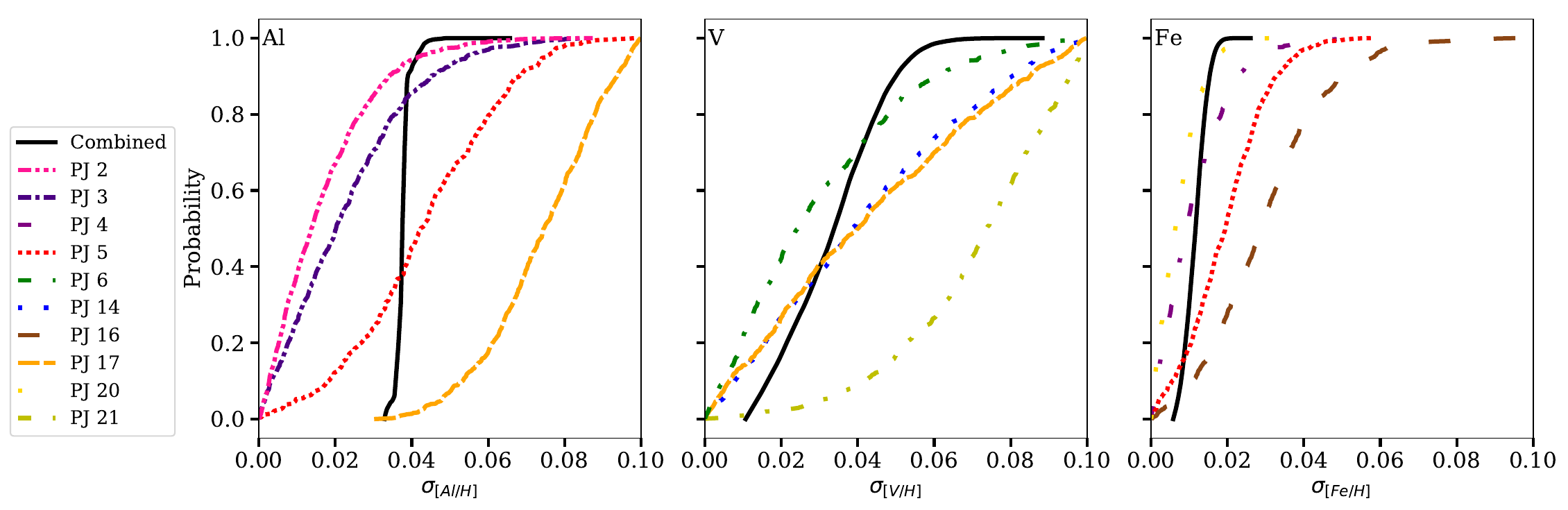}
    \caption{A summary plot of the cumulative posterior distribution functions for the intrinsic abundance scatter in Al, V, and Fe in a selection of chemically tagged birth clusters from \citet{natalie}.  In each panel, the black curve represents the cumulative posterior distribution function of the combined clusters.  The subsequent curves in each panel represent the clusters that have the maximum and minimum upper limits in each element, as well as those that have upper limits approximately 33\% and 66\% of the way between the minimum and maximum.  The 95 per cent upper limits for each PJ birth cluster are reported in Table~\ref{tab:my_results_pj}.}
    \label{fig:posterior_pj}
\end{figure*}

A summary of the cumulative posterior distribution functions for Al, V, and Fe, computed using the values of $\sigma_{\text{[X/H]}}$ captured by threshold boxes as in Figure~\ref{fig:summ_stats}, is shown in Figure~\ref{fig:posterior_pj}.  We choose to display these elements for the same reasons as above.  We show these summary curves instead of the individual cumulative posterior distribution functions for ease of visualization.  Since we find minimal evidence for abundance scatter overall (excluding those elements that have very few pixels), we can combine the cumulative posterior distribution functions to obtain a combined constraint on the intrinsic abundance scatter.  Specifically, in this calculation, we assume all of the clusters have the same intrinsic abundance scatter and we place a limit on this (B16).  This combined cumulative posterior distribution is shown by the black curve in each panel, while the four additional curves represent the cumulative posterior distribution functions for the clusters that have the maximum and minimum upper limits in each element, as well as those that have upper limits approximately 33\% and 66\% of the way between the minimum and maximum.  

From these cumulative posterior distribution functions, we can quantify the upper limits on the intrinsic abundance scatter of each birth cluster.  Table~\ref{tab:my_results_pj} describes the upper limits on the intrinsic abundance scatter at 95 per cent confidence for each of the 15 APOGEE elements for the blindly chemically tagged birth clusters we analyse.  

\begin{table*}
    \centering
    \caption{The 95 per cent upper limits on the constrained intrinsic abundance scatter for the 15 APOGEE elements in the PJ clusters that we analyse.  A dashed entry indicates that not enough data are available for the element in question after we apply the masking and uncertainty considerations.}
    \begin{tabular}{c|c|c|c|c|c|c|c|c|c}
\hline																			
	&	PJ 2	& 	PJ 3	&	PJ 4	&	PJ 5	&	PJ 6	&	PJ 9	&	PJ 10	&	PJ 11	&	PJ 12	\\
\hline																			
C	&	0.033	& 	0.017	&	0.026	&	0.043	&	0.017	&	0.049	&	0.028	&	0.026	&	0.048	\\
N	&	0.093	& 	0.059	&	0.053	&	0.047	&	0.03	&	0.075	&	0.059	&	0.067	&	0.065	\\
O	&	0.043	& 	0.063	&	0.062	&	0.068	&	0.02	&	0.034	&	0.029	&	0.041	&	0.045	\\
Na	&	0.094	& 	-	&	-	&	0.096	&	0.094	&	0.087	&	0.096	&	0.092	&	0.095	\\
Mg	&	0.04	& 	0.038	&	0.07	&	0.059	&	0.033	&	0.049	&	0.061	&	0.042	&	0.039	\\
Al	&	0.04	& 	0.055	&	0.094	&	0.088	&	0.047	&	0.085	&	0.078	&	0.073	&	0.063	\\
Si	&	0.04	& 	0.037	&	0.038	&	0.082	&	0.034	&	0.051	&	0.06	&	0.034	&	0.042	\\
S 	&	0.096	& 	0.086	&	0.089	&	0.093	&	0.084	&	0.085	&	0.079	&	0.09	&	0.078	\\
K	&	0.087	& 	0.084	&	0.093	&	0.083	&	0.084	&	0.086	&	0.084	&	0.087	&	0.095	\\
Ca	&	0.04	& 	0.034	&	0.05	&	0.061	&	0.028	&	0.067	&	0.046	&	0.05	&	0.042	\\
Ti	&	0.055	& 	0.069	&	0.088	&	0.087	&	0.042	&	0.079	&	0.06	&	0.06	&	0.075	\\
V	&	0.093	& 	0.077	&	0.093	&	0.097	&	0.073	&	0.095	&	0.089	&	0.09	&	0.093	\\
Mn	&	0.062	& 	0.03	&	0.049	&	0.059	&	0.024	&	0.035	&	0.026	&	0.035	&	0.062	\\
Fe	&	0.023	& 	0.023	&	0.027	&	0.038	&	0.022	&	0.03	&	0.028	&	0.022	&	0.028	\\
Ni	&	0.043	& 	0.04	&	0.048	&	0.078	&	0.037	&	0.063	&	0.038	&	0.042	&	0.046	\\
\hline	
    \end{tabular}
    \begin{tabular}{c|c|c|c|c|c|c|c|c|c}
\hline																			
	&	PJ 14	&	PJ 15	&	PJ 16	&	PJ 17	&	PJ 18	&	PJ 19	&	PJ 20	&	PJ 21	&	Combined	\\
\hline																			
C	&	0.024	&	0.033	&	0.059	&	0.048	&	0.032	&	0.043	&	0.031	&	0.044	&	0.019	\\
N	&	0.06	&	0.049	&	0.091	&	0.077	&	0.045	&	0.089	&	0.047	&	0.089	&	0.043	\\
O	&	0.037	&	0.051	&	0.05	&	0.088	&	0.059	&	0.044	&	0.035	&	0.083	&	0.029	\\
Na	&	0.093	&	0.095	&	0.092	&	0.092	&	-	&	0.098	&	0.093	&	0.097	&	0.085	\\
Mg	&	0.034	&	0.072	&	0.059	&	0.082	&	0.047	&	0.07	&	0.03	&	0.085	&	0.043	\\
Al	&	0.048	&	0.043	&	0.059	&	0.097	&	0.058	&	0.058	&	0.044	&	0.097	&	0.054	\\
Si	&	0.079	&	0.052	&	0.061	&	0.058	&	0.06	&	0.078	&	0.027	&	0.091	&	0.034	\\
S 	&	0.095	&	0.086	&	0.094	&	0.088	&	0.096	&	0.089	&	0.083	&	0.097	&	0.085	\\
K	&	0.092	&	0.086	&	0.093	&	0.093	&	0.087	&	0.088	&	0.085	&	0.094	&	0.085	\\
Ca	&	0.037	&	0.035	&	0.093	&	0.09	&	0.071	&	0.069	&	0.051	&	0.091	&	0.043	\\
Ti	&	0.041	&	0.049	&	0.093	&	0.092	&	0.05	&	0.068	&	0.045	&	0.085	&	0.054	\\
V	&	0.089	&	0.092	&	0.094	&	0.093	&	0.08	&	0.091	&	0.078	&	0.098	&	0.084	\\
Mn	&	0.047	&	0.05	&	0.064	&	0.048	&	0.044	&	0.038	&	0.019	&	0.075	&	0.021	\\
Fe	&	0.04	&	0.028	&	0.058	&	0.045	&	0.029	&	0.044	&	0.017	&	0.054	&	0.021	\\
Ni	&	0.046	&	0.047	&	0.064	&	0.045	&	0.041	&	0.062	&	0.031	&	0.094	&	0.04	\\
\hline																			
    \end{tabular}
    \label{tab:my_results_pj}
\end{table*}

Overall, we obtain similar or stronger limits on the intrinsic abundance scatter in the PJ clusters than the OCCAM clusters, with the combined constraints generally being strong.  Specifically, we obtain strong constraints ($\lesssim 0.05$ dex at 95 per cent confidence) on every element except for Na, Al, S, K, Ti, and V.  

It is not surprising that we obtain comparable or stronger limits on the intrinsic abundance scatter of the chemically tagged birth clusters compared to the open clusters - while the open clusters are still gravitationally bound and the birth cluster members are no longer in close proximity to each other, these results are consistent with the analysis completed by \citet{natalie}, where each chemically tagged birth cluster was found to be exceptionally chemically homogeneous when compared to a chemically similar population.  Each chemically similar population consisted of a selection of stars across APOGEE that have [Fe/H] and [Mg/Fe] within 0.05 dex of the median value of those abundances in each birth cluster.  Moreover, strong limits are to be expected in the PJ clusters, since they were identified by examining chemical abundances, and so they will be tightly clustered on these abundances, whereas the OCCAM open clusters are only spatially associated and so may not have as tight limits.  For more information on how the PJ clusters were identified and their likelihood of being true birth clusters, see \citet{natalie}.  The fact that we have achieved results consistent with the original analysis in \citet{natalie}, with regards to the question of whether or not these clusters are homogeneous, helps to strengthen the claim that these candidates are true birth clusters.  Further study of this interesting set of stars may solidify their membership, giving us more insight into the dynamical evolution of the Galactic disc.

While some of the constraints on the intrinsic abundance scatter in the PJ birth clusters are stronger than those in the example OCCAM open clusters, we discussed in Section~\ref{sec:results_summstat} that there are some signs of potential chemical inhomogeneity in some individual elements.  However, despite these factors that could be resulting in small amounts of chemical inhomogeneity, we are still able to obtain strong combined constraints on nearly every element with enough pixels available for analysis.  The fact that we see minimal levels of chemical inhomogeneity and in fact are mostly able to obtain very strong constraints is a promising test of the power of this method of constraining abundance scatter - even though the member stars of these chemically tagged birth clusters are scattered across the Galactic disc and are no longer related to each other spatially, we are still able to constrain their chemical homogeneity in a relatively robust way.

\subsection{Summary of PJ abundance scatter constraints}\label{sec:results_summaryplts}
\begin{figure}
    \centering
    \includegraphics[width=\columnwidth]{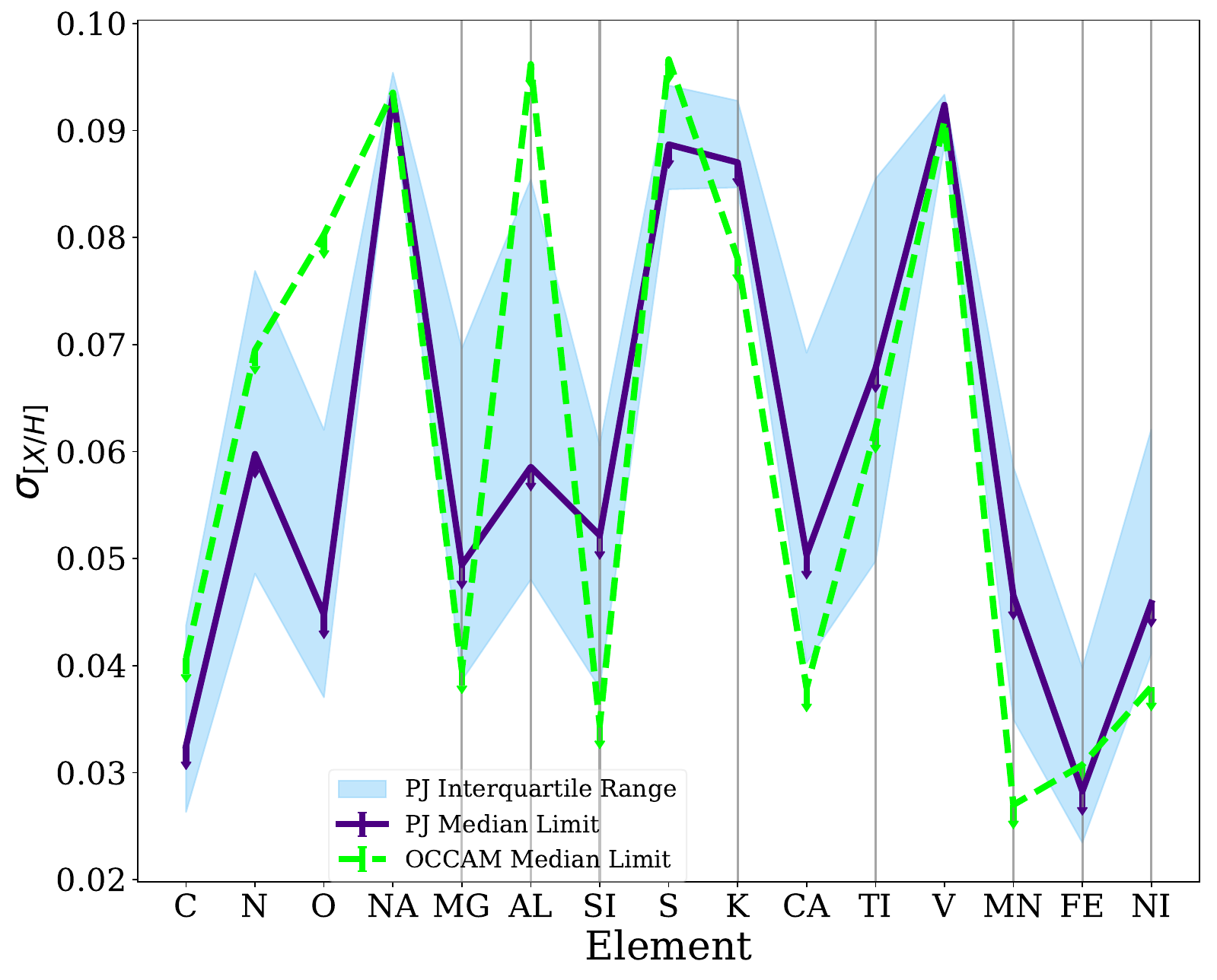}
    \caption{A summary plot of the constraints we obtain for each element X in all of the studied clusters.  The median constraint on each element across the chemically tagged birth clusters is represented by the purple arrows, connected by the solid purple line for ease of visualization.  The interquartile range is represented by the blue region.  The elements used by \citet{natalie} to group the birth clusters are highlighted by vertical grey lines.  The median constraint across the OCCAM open clusters is represented by the green arrows, connected by the dashed green line.}
    \label{fig:results_pj_elem_summary}
\end{figure}

Figure~\ref{fig:results_pj_elem_summary} provides a visual summary of the constraints we obtain for each element in the PJ clusters (as well as the OCCAM open clusters, for comparison).  The median constraint on each element across all of the birth clusters is represented by the purple arrows, the median constraint across the open clusters is represented by the green arrows, and the interquartile range across the birth clusters is represented by the blue region.  The arrows for each set of clusters have been connected by lines for ease of visualization.  Elements used by \citet{natalie} to associate the birth clusters are highlighted using vertical grey lines (i.e. Mg, Al, Si, K, Ti, Mn, Fe, and Ni).  

\begin{figure*}
    \centering
    \includegraphics[width=\textwidth]{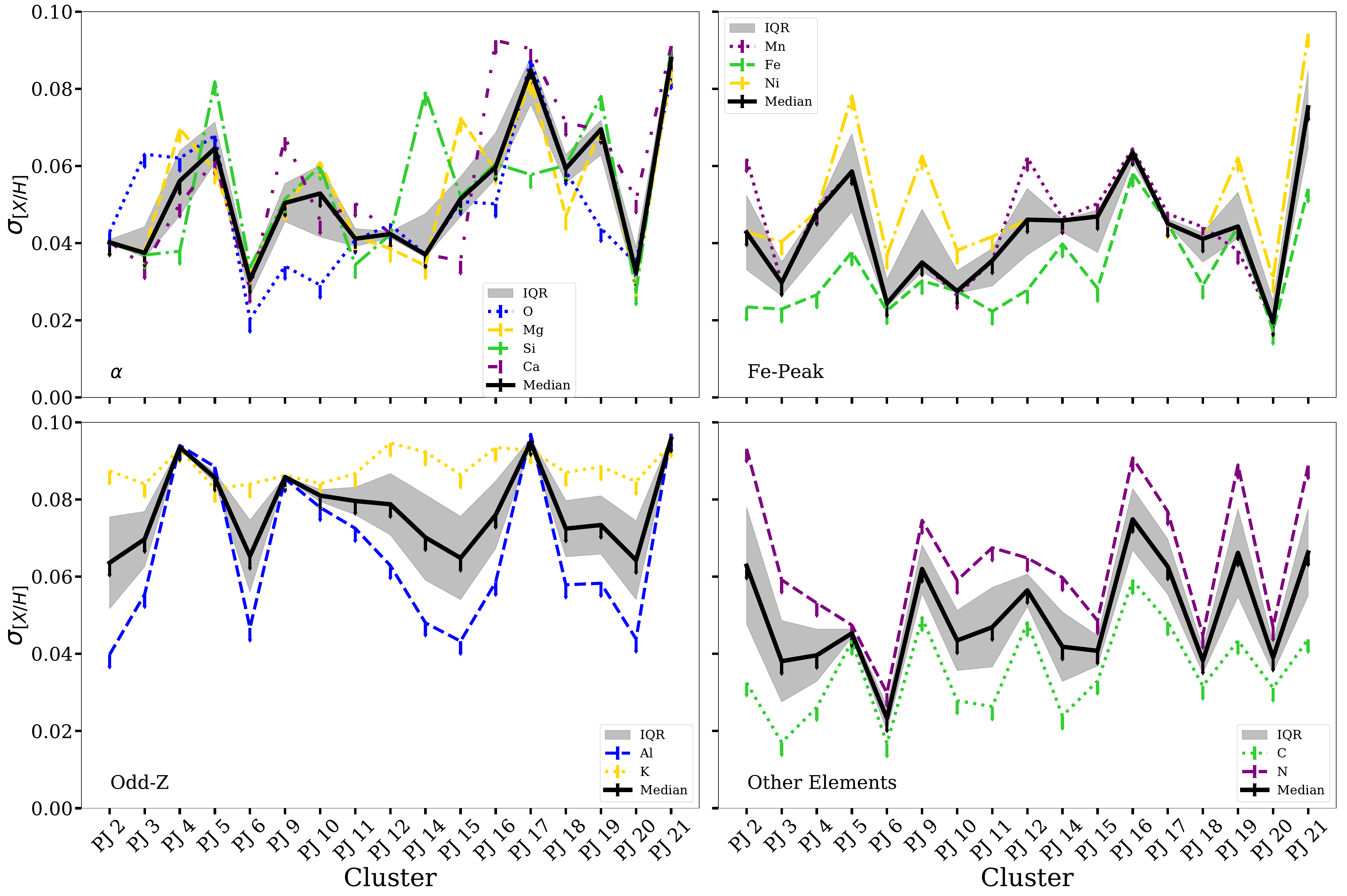}
    \caption{A summary plot of the constraints that we obtain for each chemically tagged birth cluster, broken up into different elemental groupings.  The constraints on some of the $\alpha$-elements are shown in the top left panel, those on the Fe-peak elements are shown in the top right panel, those on some of the odd-Z elements are shown in the bottom left panel, and those on C and N are shown in the bottom right panel.  In each panel, the medians of the constraints for the elements shown are represented by the black arrows, connected by the solid black line, and the interquartile ranges are represented by the grey region.  We exclude some of the elements with few pixels available for analysis from this figure (i.e. Na, S, Ti, and V).}
    \label{fig:results_pj_cluster_summary}
\end{figure*}

Figure~\ref{fig:results_pj_cluster_summary} provides a visual summary of our constraints across the PJ clusters.  We have grouped these results into $\alpha$-elements, Fe-peak elements, odd-Z elements, and C and N.  We exclude some of those elements with few pixels available for analysis from this visualization (i.e. Na, S, Ti, and V), as their constraints are generally weak and not representative of the entire sample.  The medians of the constraints for the elements shown in each panel are represented by the black arrows, connected by the solid black lines for ease of visualization.  The interquartile ranges of these constraints are represented by the grey regions.

It is clear from Figure~\ref{fig:results_pj_elem_summary} that the constraints in each element in the PJ clusters are generally well below $0.1$ dex, excluding those elements that do not have many pixels available for analysis.  The constraints for C and Fe are the strongest.  Similar conclusions can be made by examining Figure~\ref{fig:results_pj_cluster_summary} for various constraints in each birth cluster.  In excluding most of the elements with few pixels available for analysis from these figures, it is clear that most of the elements in the majority of the birth clusters have very strong constraints.

\section{Implications for chemical tagging}\label{sec:discussion}
The results of B16 held promise for chemical tagging as a result of the fact that tight constraints were found for each cluster studied.  The somewhat disparate results of this study, with a few elements in some clusters displaying intrinsic abundance scatter, may result in different implications for chemical tagging studies.  

B16 cites several advantages and disadvantages for the method we use compared to the traditional method of determining chemical homogeneity in open clusters through derived abundances.  These advantages and disadvantages and their implications for chemical tagging certainly hold for several of the PJ clusters and are worth discussing here.  First, while there are some sources of systematic uncertainty that are difficult to account for with this method, it still allows us to avoid some common complications.  For example, in traditional studies, measuring consistent abundances across various stellar types can be challenging, due to ambiguities in the elemental line list, one-dimensional radiative transfer, and inconsistencies with LTE assumptions \citep{Ness_2015}.  In our method, however, these uncertainties can be modelled as smooth functions of initial stellar mass, and therefore $T_{\text{eff}}$, which is the foundation of our model.  Additionally, since this method does not prescribe any behaviour to the overall $T_{\text{eff}}$ trends of the stellar spectra used, we directly avoid the uncertainties stemming from all of these effects.  Another common problem for chemical tagging studies is the inhomogeneity of present-day chemical abundances in cluster members due to stellar-evolutionary effects on surface abundances (i.e. deep mixing, gravitational settling, or atomic diffusion, \citealt{Dotter}), which are difficult to model.  However, the consequences of these effects are predominantly deterministic functions of stellar mass, so our modelling of stellar spectra as a one-dimensional function of initial stellar mass allows us to constrain the \textit{initial} abundance scatter in the chemically tagged birth clusters, as opposed to the present-day scatter (B16).  Finally, since this technique makes use of forward simulations and ABC, realistic complications seen in the observed spectra can be taken into account in these simulations (i.e. correlated spectral noise and uncertainties that result from the applied continuum normalization, B16).

It is true that we find indication of chemical inhomogeneity in some of the elements in a few of the PJ clusters (summarized in Table~\ref{tab:bad_elems}).  If this is a real effect, this potential evidence of chemical inhomogeneity may not spell the end for chemical tagging studies.  Previously, it was thought that chemical homogeneity was a requirement for chemical tagging \citep{Freeman_2002}.  However, as we discussed in Section~\ref{sec:introduction}, \citet{casamiquela} studied the Hyades, Praesepe, and Rupecht 147 open clusters.  They found that there was a certain level of chemical inhomogeneity in each of these clusters, however, it was still possible to distinguish the chemical signature of Ruprecht 147 from the Hyades and Praesepe.  Perhaps this is an indication that the requirements for chemical tagging can be relaxed - while chemical homogeneity is required for chemical tagging, the level of homogeneity may not need to be as strong as previously thought.  Additionally, these inhomogeneous results still allow for a greater understanding of the chemical and dynamical history of the Galactic disc.  If clusters like PJ 5, for example, are truly inhomogeneous, this may indicate a lack of the uniformity that is usually assumed in the early stages of the formation of the Galaxy.

Our ability to strongly constrain the intrinsic abundance scatter in the majority of the chemically tagged birth clusters from \citet{natalie} indicates great promise for chemical tagging studies.  As stated in Section~\ref{sec:introduction}, the ability to reconstruct star clusters that have been dispersed across the Galactic disc will reveal crucial information about the chemical and dynamical evolution of the Galactic disc.  By constraining the levels of chemical homogeneity within reconstructed star clusters, we are able to have a greater understanding of the conditions under which these groups and the stars within these groups formed.

Additionally, in finding that the levels of chemical homogeneity in these chemically tagged birth clusters are similar to or stronger than those in various open clusters, consistent with findings in \citet{natalie}, we have further solidified the association of these chemically tagged stars as members of the same birth clusters.  This is encouraging for future chemical tagging studies and can also allow us to study these specific groups in more detail.  For example, we can more closely study the migration patterns of these stars as they separated from their common formation sites and travelled across the Galactic disc.

\section{Summary and conclusions}\label{sec:conclusion}
In summary, we constrain the initial abundance spread in several blindly chemically tagged birth clusters identified in \citet{natalie}.  We determine these constraints using a method first developed by B16, whereby we model the stellar spectra as a one-dimensional function of initial stellar mass, perform quadratic fits as a function of T$_{\text{eff}}$ as a proxy for stellar mass, and produce cumulative distributions of the uncertainty-normalized fit residuals.  We complete forward modelling of the spectra by generating synthetic APOGEE spectra using the \textsc{psm} code \citep{psm} and a variety of values of intrinsic abundance scatter, drawn from a uniform prior of $0$ to $0.1$ dex.  We compare the synthetic spectra to the observed spectra via ABC through the use of two summary statistics: the Kolmogorov-Smirnov distance ($D_n$) and a covariance matrix statistic ($|\Delta\text{Cov}_{\text{ij}}|$).  We produce cumulative posterior distribution functions of the values of intrinsic abundance scatter of the simulations that are the closest to the data and we obtain upper limits on the intrinsic abundance scatter at 95 per cent confidence.  We exemplify this method using the open clusters M67, NGC 6819, and NGC 6791, data for which we obtain from the Open Cluster Chemical Analysis and Mapping survey \citep{Donor_2018}.  For M67 and NGC 6819, which were also studied in B16, we obtain similar results on the intrinsic abundance scatter of these clusters, within uncertainties introduced by differences in our methods.

In general, we find very strong limits on the intrinsic abundance scatter of the chemically tagged birth clusters, with strong constraints ($\lesssim 0.5$ dex at 95 per cent confidence) on the abundance scatter in every element except for those which have very few pixels available for analysis.  While we see some evidence for a small amount of chemical inhomogeneity in some of the elements in a few of the PJ birth clusters, we are still able to obtain similar or stronger combined limits on the intrinsic abundance scatter of these clusters compared to the OCCAM open clusters, consistent with findings in \citet{natalie}.  

These results hold great promise for the method of chemical tagging.  We find that most of the elements in most of the chemically tagged birth clusters have strong levels of chemical homogeneity.  For the few that show some intrinsic abundance scatter, it may be possible to relax the constraints on the level of chemical homogeneity that is required for chemical tagging and distinguish the chemical signatures of different stellar groups, as found by \citet{casamiquela}.  Regardless, by strongly constraining the levels of chemical homogeneity within reconstructed stellar birth clusters, we are able to further solidify the association of these chemically tagged stars as birth clusters and to have a clearer understanding of the chemical and dynamical history of the Galactic disc.

\section*{Data availability}

The data upon which this research is based are publicly available as part of SDSS-IV's DR14 data release (\url{https://www.sdss.org/dr14}).

\section*{Acknowledgements}

We acknowledge and are grateful to have the opportunity to work on the land on which the University of Toronto operates.  For thousands of years it has been the traditional land of the Huron-Wendat, the Seneca, and the Mississaugas of the Credit River, and it is home to many Indigenous peoples from across Turtle Island.  We also acknowledge and are grateful to have access to the data collected on the land on which the Apache Point Observatory is located, the traditional land of the Pueblo of Sandia, the Navajo, and the Apache. 

We thank the anonymous referees for a useful report.  We also thank John Donor, Vijith J. Poovelil, and our other colleagues in SDSS for helpful comments and conversations that enhanced this work.  This research received financial support from NSERC (funding reference numbers RGPIN-2015-05235 \& RGPIN-2020-04712), an Ontario Early Researcher Award (ER16-12-061), and from the Canada Research Chair program. NPJ received support from an Alexander Graham Bell Canada Graduate Scholarship-Doctoral from the Natural Sciences and Engineering Research Council of Canada.

Funding for the Sloan Digital Sky Survey IV has been provided by the Alfred P. Sloan Foundation, the U.S. Department of Energy Office of Science, and the Participating Institutions. SDSS-IV acknowledges support and resources from the Center for High-Performance Computing at the University of Utah. The SDSS web site is \url{www.sdss.org}.



\bibliographystyle{mnras}
\bibliography{chem_homogeneity} 


\bsp	
\label{lastpage}
\end{document}